\documentclass[preprint, showpacs,preprintnumbers,amsmath,amssymb,nofootinbib]{revtex4}
\usepackage{amssymb}
\usepackage{natbib}
\usepackage{graphicx}
\usepackage{dcolumn}
\usepackage{bm}

\newcommand{\bea}{\begin{eqnarray}}
\newcommand{\eea}{\end{eqnarray}}
\begin{document}
\title{ Testing the FLRW metric with the Hubble and transversal BAO measurements}
\author{Min Wang$^1$, Xiangyun Fu$^1$\footnote{corresponding author: xyfu@hnust.edu.cn}, 
Bing Xu$^2$\footnote{corresponding author: xub@ahstu.edu.cn},
Ying Yang$^1$, and Zhaoxia Chen$^1$}
\address{ $^1$Department of Physics, Key Laboratory of Intelligent Sensors and Advanced Sensor Materials,\\Hunan University of Science and Technology, Xiangtan, Hunan 411201, China\\
$^2$School of Electrical and Electronic Engineering, Anhui Science and Technology University, Bengbu, Anhui 233030, China}
\begin{abstract}
The cosmological principle is one of the fundamental assumptions of the standard model of Cosmology (SCM), and it allows us to describe cosmic distances and clocks by using the Friedmann-Lema$\rm{\hat{{\i}}}$tre-Roberton-Walker (FLRW) metric. Thus, it is essential to test the FLRW metric with cosmological observations to verify the validity of the SCM. In this work, we perform tests of the FLRW metric by comparing the observational comoving angles between the Hubble $H(z)$ and the transversal Baryon Acoustic Oscillation (BAO) measurements. The Gaussian process is employed to reconstruct the Hubble $H(z)$ measurements and the angular diameter distance (ADD) from the transversal BAO data.  A non-parametric method is adopted to probe the possible deviations from the FLRW metric at any redshift by comparing the comoving distances from the reconstructed  Hubble $H(z)$ measurements with the ADD reconstructed from the transversal BAO data. Then, we propose  two types of parameterizations for the deviations from the FLRW metric, and  test the FLRW metric by using the priors of specific sound horizon scales.  To avoid the bias caused by the prior of a specific sound horizon scale, we perform the consistency test with a flat prior of the sound horizon scale. We find that there is a concordance between the FLRW metric and the observational data by using parametric and non-parametric methods, and the parameterizations can be employed to test the FLRW metric in a new way independent of the sound horizon scale.
 $\mathbf{Keywords:}$  Cosmological principle, FLRW metric, parametric method
\end{abstract}
 \maketitle
\section{Introduction}
In 1998, the unexpected dimming of type Ia supernova (SNIa) revealed the evidence of the accelerating expansion of the Universe for the first time~\cite{Riess1998,Perlmutter1999}. In the frame of
General Relativity, a cosmic distribution of an exotic component with negative pressure, dubbed as dark energy, has
been suggested to explain the accelerating expansion. In addition, a type of cold dark matter in the Universe is needed to explain the formation of the cosmic structure  and galaxy dynamics. So far, the cosmological constant $\Lambda$ is the best candidate to explain the so-called dark energy. Thus, the standard model of Cosmology (SCM), a flat cosmological constant $\Lambda$ Cold
Dark Matter ($\Lambda$CDM) model, has been established successfully to explain cosmological observations from the
Cosmic Microwave Background (CMB)~\cite{Aghanim2020}, SNIa luminosity distances~\cite{Scolnic2018}, as well as the clustering and weak lensing of the large-scale structure~\cite{Alam2021,Heymans2021,Abbott2022,Secco2022}.
The SCM scenario is based upon two fundamental assumptions. One is that General Relativity can be considered as the underlying theory of gravity. The other is that the Universe is statistical homogeneity and isotropy at large scales, also known as the Cosmological Principle (CP). In modern cosmology, the CP is formulated as follows: at suitably large scales, the average evolution of the Universe is exactly governed by the Friedmann-Lema$\hat{\rm{\i}}$tre-Robertson-Walker (FLRW) metric~\cite{Aluri2023}.
The homogeneous $\Lambda$CDM model is a highly successful model that is  compatible with almost all observations up to now. However, the SCM faces the challenges of the Hubble tension~\cite{Krishnan2021}, fine-tuning problem~\cite{Marks2022}, and cosmic coincidence problems~\cite{Weinberg1989,Sahni2000}. The Hubble tension comes from  the discrepancy between local measurements
based on Cepheids and SNIa~\cite{Riess2019} and
analysis of the CMB~\cite{Aghanim2020} in the framework of the $\Lambda$CDM model. These problems  imply that there is a potential departure from the
SCM,  and the
$\Lambda$CDM model needs further adjustments~\cite{Efstathiou}.  An effective way to support or exclude the $\Lambda$CDM paradigm is to verify  the validity of the FLRW metric through new methods and  various astronomic observations.

Maartens first proposed that homogeneity means a consistency relation  between the angular diameter distances and comoving distances in the FLRW universe~\cite{Maartens2011}. Any violation of the consistency relation implies
a non-FLRW universe. Therefore, the consistency
relations between cosmic distances and chronometers provide us with an opportunity to test the FLRW metric. Recently,  Arjona and Nesseris introduced a function $\zeta^\prime(z)=1-{\theta_{H(z)}/\theta_{\rm BAO}}$ to define the possible deviations from the FLRW metric, and tested  cosmic homogeneity by comparing the Hubble $H(z)$ measurements from the clustering and radial Baryon Acoustic Oscillation (BAO) data with the transversal BAO measurements~\cite{Arjona2021}. Here, the ${\theta_{H(z)}}$ and ${\theta_{\rm BAO}}$ denote comoving angles from the Hubble $H(z)$ and transversal BAO measurements, respectively.  The value of function $\zeta^\prime(z)$ at any redshift was obtained with the method of genetic algorithms. Using the sound horizon scale based on $\Lambda$CDM model, they found that both
reconstructions are consistent with the flat $\Lambda$CDM
model at the $1\sigma$ confidence level (CL) and at $2\sigma$ CL, respectively.  Then, Bengaly reconstructed the function $\zeta^\prime(z)$ with
the Gaussian process and  tested the consistency relation by comparing the 18 radial BAO data points with
transversal BAO data~\cite{Bengaly2022}. A mild deviation of the FLRW metric was obtained
at $3.5\sigma$ CL in the relatively low redshift range $0.1<z<0.3$ under the priors of the SH0ES $H_0$ and the specific sound horizon scales $r_{\rm s}$ from the Sloan Digital Sky Survey (SDSS) data release 11 galaxies~\cite{Carvalho2020} and BAO measurements~\cite{Verde2017}.

It is easy to see that both tests in Refs.~\cite{Arjona2021,Bengaly2022} are dependent on the values of the present Hubble parameter $H_0$ and the sound horizon scale $r_{\rm s}$. It should be noted that the so-called fitting problem remains a challenge for BAO peak location as a standard ruler~\cite{Ellis1987}, although the BAO
measurements are employed to analyze various cosmological parameters. In particular, Roukema {\it et al.} recently detected the environmental dependence of BAO location~\cite{Roukema2015,Roukema2016}. Moreover, Ding {\it et al.} and Zheng {\it et al}. pointed out a noticeable systematic difference between Hubble $H(z)$ measurements based on BAO and those obtained with differential aging techniques~\cite{Ding2015,Zheng2016}. In addition, different sound horizon scales $r_{\rm s}$ and the present Hubble parameter $H_0$ are obtained from various observational data, such as, CMB observations~\cite{Ade2016,Bennett2013},  the SDSS data release 11 galaxies~\cite{Carvalho2020}, and BAO measurements~\cite{Verde2017}. Any prior of $r_{\rm s}$ may bring bias on the test of the FLRW metric.  Therefore, it is meaningful to test the FLRW metric with new methods independent of the sound horizon scales $r_{\rm s}$, which is the main motivation of this work.

We first test the FLRW metric by comparing the observational comoving angles from the Hubble $H(z)$ measurements with that from the transversal BAO data. The Hubble parameter $H(z)$ measurements are either obtained by the differential age method  or by the determination of the
BAO peak in the radial direction (hereafter referred to as radial BAO observations). The function $\zeta(z)=1-\theta_{\rm BAO}/{\theta_{H(z)}}$ is employed to quantify the possible deviation from the FLRW metric.  To obtain the observed value of function $\zeta(z)$ at any redshift, the Gaussian process is employed to reconstruct the Hubble $H(z)$ measurements and the angular distances from the transversal BAO data. The results show that the best-fit values of function $\zeta(z)$ vary with redshift $z$.
Then, we propose two general types of parametrizations for the function $\zeta(z)$,  namely, P1: $\zeta(z)=\zeta_0 (1+z)$ and P2:  $\zeta(z)=\zeta_0/(1+z)$. Here, the non-zero constant $\zeta_0$ represents a possible deviation from the FLRW metric. We test the FLRW metric successively by using the priors of specific sound horizon scales $r_{\rm s}$ and a flat priors of $r_{\rm s}$. We show that the FLRW metric is consistent with the observational data, and the parametric method provides an effective way independent of the sound horizon scale $r_{\rm s}$ to test the consistency relation.

\section{Data and Methodology}
\subsection{observational data}
The Hubble parameter $H(z)$ measurements in our analysis can be obtained by two interrelated methods. The first compilation  comes from the differential age (DA) method proposed in Ref.~\cite{Jimenez2002}, where the ages of early-type galaxies are compared with the same metallicity and separated by small redshift intervals. In General Relativity, the Hubble parameter $H(z)$ can be also written as
\begin{equation}\label{hz}
H(z)=-{1\over {1+z}}{dz\over dt}\,,
\end{equation}
where $dz/dt$ is measured using the $4000{\rm {\AA}} $ break feature as function of redshifts.
Thus, this approach directly measures the Hubble parameter by using spectroscopic dating of passive evolutionary galaxies to compare their ages and metallicities, providing $H(z)$ measurements in a cosmological model-independent method~\cite{Zhang2014,Stern2010,Moresco2012,Ratsimbazafy2017,Moresco2016,Moresco2015,Yang2023,Marra2018,Wu2007}.  Here, the 31 cosmological model-independent
data points are compiled in Table~\ref{Hub31}. The second $H(z)$ compilation
comes from the clustering of galaxies or quasars, being a
direct probe of the Hubble expansion by determining the
BAO peak in the radial BAO observations~\cite{Gaztanaga2009,Magana2018}. $18$ Hubble $H(z)$ measurements of the radial BAO mode are compiled in Table~\ref{HubRad18}. It should be noted that the observed Hubble data from radial BAO mode are obtained with an underlying $\Lambda$CDM cosmological model. So, any method using Hubble data obtained from the radial BAO measurements  is not completely cosmological model-independent.
\begin{table*}[htb]
\label{tab:results}
\begin{center}
\begin{tabular}{ccccc|cccc|ccccc} \hline \hline
& $z$ & $H(z)$ & $\sigma_{ H(z)}$ &  Ref. & $z$ & $H(z)$ & $\sigma_{ H(z)}$ &  Ref.  & $z$ & $H(z)$ & $\sigma_{ H(z)}$ &  Ref. \\  [1ex] \hline
& 0.07 & 69 & 19.6 & ~\cite{Zhang2014} & 0.4 & 95 & 17 & ~\cite{Stern2010} & 0.88 & 90 & 40 & ~\cite{Stern2010} \\ [1ex]
& 0.10  & 69 & 12 & ~\cite{Stern2010}  & 0.4004 & 77 & 10.2 & ~\cite{Moresco2016}  & 0.9  & 117 & 23 & ~\cite{Stern2010}     \\ [1ex]
& 0.12 & 68.6 & 26.2 & ~\cite{Zhang2014}  & 0.4247  & 87.1 & 11.2 & ~\cite{Moresco2016}  & 1.037 & 154 & 20 & ~\cite{Moresco2012}     \\ [1ex]
& 0.17 & 83 &  8 & ~\cite{Stern2010} & 0.4497 & 92.8 & 12.9 & ~\cite{Moresco2016} & 1.3 & 168 &  17 & ~\cite{Stern2010}    \\ [1ex]
& 0.1791 & 75 &  4 & ~\cite{Moresco2012} & 0.47 & 89 & 34 & ~\cite{Ratsimbazafy2017} & 1.363 & 160 &  33.6 & ~\cite{Moresco2015}    \\ [1ex]
& 0.1993 & 75 &  5 & ~\cite{Moresco2012} & 0.4783 & 80.9 & 9 & ~\cite{Moresco2016} & 1.43 & 177 &  18 & ~\cite{Stern2010}    \\ [1ex]
& 0.20 & 72.9 &  29.6 & ~\cite{Zhang2014} & 0.48 & 97 & 62 & ~\cite{Stern2010} & 1.53 & 140 &  14 & ~\cite{Stern2010}    \\ [1ex]
& 0.27 & 77 &  14 & ~\cite{Stern2010} & 0.5929 & 104 & 13 & ~\cite{Moresco2012} & 1.75 & 202 &  40 & ~\cite{Stern2010}    \\ [1ex]
& 0.28 & 88.8 &  36.6 & ~\cite{Zhang2014} & 0.6797 & 92 & 8 & ~\cite{Moresco2012} & 1.965 & 168.5 &  50.4 & ~\cite{Moresco2015}    \\ [1ex]
& 0.3519 & 83 &  14 & ~\cite{Moresco2012} & 0.7812 & 105 & 12 & ~\cite{Moresco2012}    \\ [1ex]
& 0.3802 & 83 & 13.5 & ~\cite{Moresco2016} & 0.8754 & 125 &  17 & ~\cite{Moresco2012}   \\ [-0.25ex]
\hline\hline
\end{tabular}
\caption[]{31 Hubble parameter measurements $H(z)$ obtained from the DA method (in units of ${\rm km \,s^{-1}Mpc^{-1}}$). }
\label{Hub31}
\end{center}
\end{table*}
\begin{table*}[htb]
\label{tab:results}
\begin{center}
\begin{tabular}{ccccc|cccc|ccccc} \hline \hline
& $z$ & $H(z)$ & $\sigma_{ H(z)}$ &  Ref. & $z$ & $H(z)$ & $\sigma_{ H(z)}$ &  Ref.  & $z$ & $H(z)$ & $\sigma_{ H(z)}$ &  Ref. \\  [1ex] \hline
& 0.24 & 79.69 & 2.65 & ~\cite{Gaztanaga2009} & 0.43 & 86.45 & 3.68 & ~\cite{Gaztanaga2009} & 0.60 & 87.9 & 6.10 & ~\cite{Blake2012} \\ [1ex]
& 0.30  & 81.70 & 6.22 & ~\cite{Oka2014}  & 0.44 & 82.60 & 7.80 & ~\cite{Blake2012}  & 0.61  & 97.3 & 2.10 & ~\cite{Alam2017}     \\ [1ex]
& 0.31 & 78.17 & 4.74 & ~\cite{Wang2017}  & 0.51  & 90.40 & 1.90 & ~\cite{Alam2017}  & 0.64 & 98.82 & 2.99 & ~\cite{Wang2017}     \\ [1ex]
& 0.35 & 82.70 &  8.40 & ~\cite{Chuang2013} & 0.52 & 94.35 & 2.65 & ~\cite{Wang2017} & 2.33 & 224.00 &  8.00 & ~\cite{Bautista2017}    \\ [1ex]
& 0.36 & 79.93 &  3.39 & ~\cite{Wang2017} & 0.56 & 93.33 & 2.32 & ~\cite{Wang2017} & 2.34 & 223.00 &  7.00 & ~\cite{Delubac2015}    \\ [1ex]
& 0.38 & 81.50 & 1.90 & ~\cite{Alam2017} & 0.57 & 98.48 &  3.19 & ~\cite{Anderson2014} & 2.36 & 227.00 & 8.00 & ~\cite{Font2014}  \\ [-0.25ex]
\hline\hline
\end{tabular}
\caption[]{18 Hubble parameter measurements $H(z)$ obtained from the radial BAO measurements (in units of ${\rm km \, s^{-1}Mpc^{-1}}$). }
\label{HubRad18}
\end{center}
\end{table*}

The 15 transversal BAO measurements~\cite{Carvalho2016,Alcaniz2017,Carvalho2018,Carvalho2020,Nunes2020} are compiled in Table~\ref{AngularBAO}, which were obtained using public
data releases of the SDSS~\cite{York2000}.  These measurements of the BAO scale can be obtained by using the angular $2$-point correlation function, which involves only the angular separation $\theta$ between pairs, yielding information of the angular diameter distance (ADD) almost model-independently, provided that the comoving sound horizons is known~\cite{Nunes2020}. The distributions of the transversal BAO measurements and Hubble $H(z)$ data are shown in Fig.~\ref{fighubble}.
\begin{table*}[htb]
\label{tab:results}
\begin{center}
\begin{tabular}{ccccc|cccc|ccccc} \hline \hline
& $z$ & $\theta_{\rm BAO}$ & $\sigma_{\rm BAO}$ &  Ref. & $z$ & $\theta_{\rm BAO}$ & $\sigma_{\rm BAO}$ &  Ref.  & $z$ & $\theta_{\rm BAO}$ & $\sigma_{\rm BAO}$ &  Ref. \\  [1ex] \hline
& 0.11 & 19.8 & 3.26 & ~\cite{Carvalho2021} & 0.49 & 4.99 & 0.21 & ~\cite{Carvalho2016} & 0.59 & 4.39 & 0.33 & ~\cite{Carvalho2020} \\ [1ex]
& 0.235  & 9.06 & 0.23 & ~\cite{Alcaniz2017}  & 0.51 & 4.81 & 0.17 & ~\cite{Carvalho2016}  & 0.61  & 3.85 & 0.31 & ~\cite{Carvalho2020}     \\ [1ex]
& 0.365 & 6.33 & 0.22 & ~\cite{Alcaniz2017}  & 0.53  & 4.29 & 0.30 & ~\cite{Carvalho2016}  & 0.63 & 3.90 & 0.43 & ~\cite{Carvalho2020}     \\ [1ex]
& 0.45 & 4.77 &  0.17 & ~\cite{Carvalho2016} & 0.55 & 4.25 & 0.25 & ~\cite{Carvalho2016} & 0.65 & 3.55 &  0.16 & ~\cite{Carvalho2020}    \\ [1ex]
& 0.47 & 5.02 & 0.25 & ~\cite{Carvalho2016} & 0.57 & 4.59 &  0.36 & ~\cite{Carvalho2020} & 2.225 & 1.77 & 0.31 & ~\cite{Carvalho2018}  \\ [-0.25ex]
\hline\hline
\end{tabular}
\caption[]{15 transversal BAO measurements $\theta_{\rm BAO}$ (in deg) and their errors $\sigma_{\rm BAO}$ at redshift $z$. These data were obtained through public data publication using SDSS. }
\label{AngularBAO}
\end{center}
\end{table*}

\subsection{Gaussian Process}
To test the consistency relation between the ADDs and comoving distances in the FLRW universe~\cite{Maartens2011}, the simplest way is to make the comparison between the radial  and transversal angular mode  at the same redshift~\cite{Maartens2011,Arjona2021}.
In principle, given a comoving observed angle mode from the Hubble measurements, one should select the corresponding one from the transversal BAO data at the same redshift $z$ to test the FLRW metric. However, this condition
usually cannot be met in recent astronomical observations.

The Gaussian process provides a powerful tool for the distribution of functions in a stochastic statistical process, and it  has gained much attention in cosmology due to its ability to reconstruct
cosmological data in a model-independent manner.
The distribution of function value is a Gaussian at each point $x$, and the
reconstruction consists of a mean function with Gaussian error bands. The function values at different points are correlated by the covariance kernel $k(x,x')$,  which depends on
the hyper-parameters of  $\sigma_f$ and $l$.  The parameter $l$ represents roughly to the distance one needs
to be moved in the input space before the function value changes
significantly, and $\sigma_f$ denotes typical changes in the
function value~\cite{Seikel20122}.  The hyper-parameters play important role in determining the
error bars of observation data. Both of the hyper-parameters
are optimized by the Gaussian process with the observed data set.
 Different Gaussian process  kernels  have been used for the null test in Ref.~\cite{Bengaly2022}, and the results showed that  different choices for kernels of Gaussian process  provided slightly more degenerate reconstruction which reduced the statistical significance of these deviations.  The Gaussian process has been employed to make constraints on cosmological constants with SNIa observation~\cite{Yahya2014}, to probe the micro-motions of cosmic matter~\cite{Gonzalez2016}, to investigate the deceleration parameter and the duality-distance parameter~\cite{Costa2015}, to constrain cosmological mixing parameters~\cite{Mukherjee2021,holanda2013}, and to infer the Hubble constant~\cite{Verde2014,Li2016,Busti2014}. More recently, Zhang, {\it et al.} used Bayes factors to  evaluate the differences between different kernel functions by analyzing the cosmic Chronometer data, SNIa, and Gamma Ray Burst. The results showed that Bayes factors indicate no significant dependence of the data on each kernel~\cite{Zhang2023}.

To employ all of 15 transversal BAO measurements to probe the deviation from the FLRW metric,
we reconstruct  the Hubble measurements with the Gaussian process~\cite{Carlos2020,Shafieloo2012} to obtain the continuous $H(z)$ function. Therefore, for each transversal BAO data point, we can obtain the corresponding Hubble measurement
with the same redshift from the reconstructed Hubble $H(z)$ function.  In our analysis, following the results obtained from Ref.~\cite{Zhang2023}, we do not consider the impact of different kernels on the consistency relation, and adopt the general covariance kernel, namely the squared exponent~\cite{Seikel2012}
\begin{equation}
k(x,x')={\sigma^{2}_f{\rm exp}{(-{{(x-x')^2}\over{2l^{2}}})}}\,.
\end{equation}

\subsection{Methodology}{\label{subsec}}
Following the approach to test the consistency relation as proposed in
Refs.~\cite{Maartens2011,Arjona2021}, deviations from the FLRW metric can be obtained by reconstructing the  comoving  distances ($D_{\rm C}$) from the Hubble $H(z)$ measurements and the ADD $D_{\rm A}$ from the transversal BAO observations. The Hubble $H(z)$ measurements in our analysis are obtained from the DA method and the radial BAO measurements, and they are listed in Table~\ref{Hub31} and Table~\ref{HubRad18}, respectively.
The comoving observed angle from the $H(z)$ data is given by
\begin{equation}
\theta_{H(z)}={r_{\rm s}\over D_{\rm C}(z)}\,,
\end{equation}
and the same for the transversal BAO data,
\begin{equation}
\theta_{\rm BAO}={r_{\rm s}\over{(1+z)}D_{\rm A}(z)}\,,
\end{equation}
where $r_{\rm s}$ is the sound horizon scale at the drag epoch, and the comoving distance $D_{\rm C}$ can be written as \begin{equation}
D_{\rm C}(z)=c\int_{0}^{z}{dz'\over{H(z)}}\,.
\end{equation}

\begin{figure}[htbp]
\includegraphics[width=7cm]{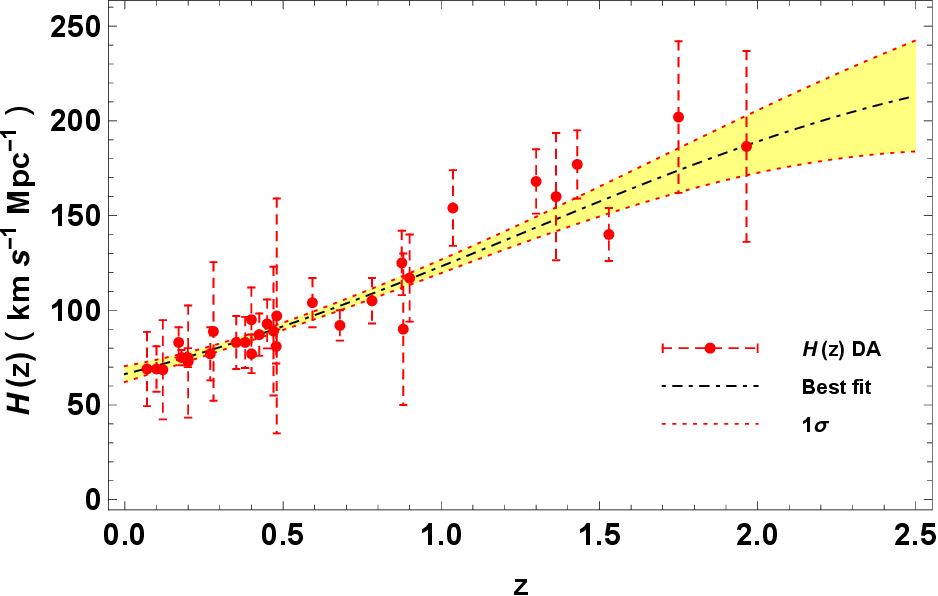}
\includegraphics[width=7cm]{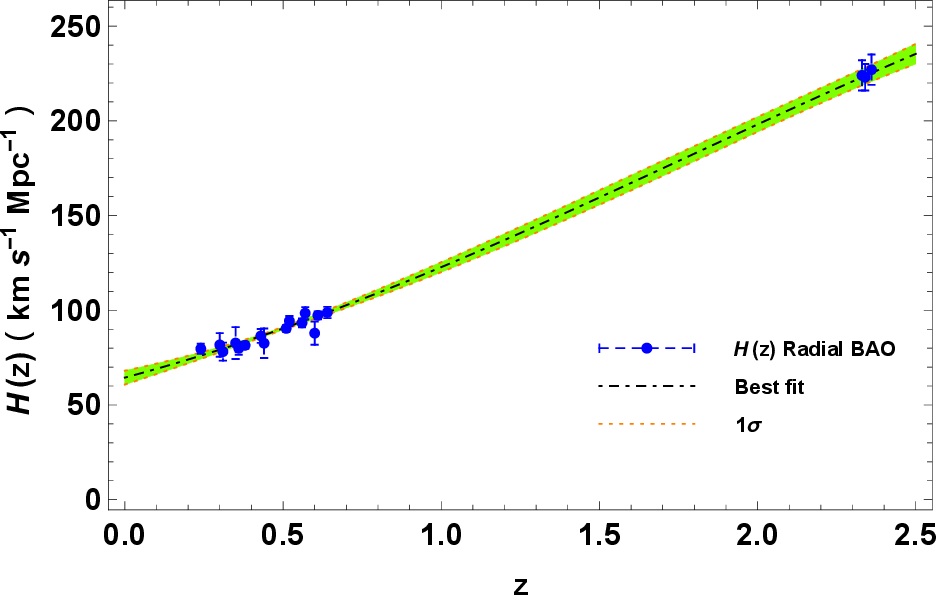}
\includegraphics[width=7cm]{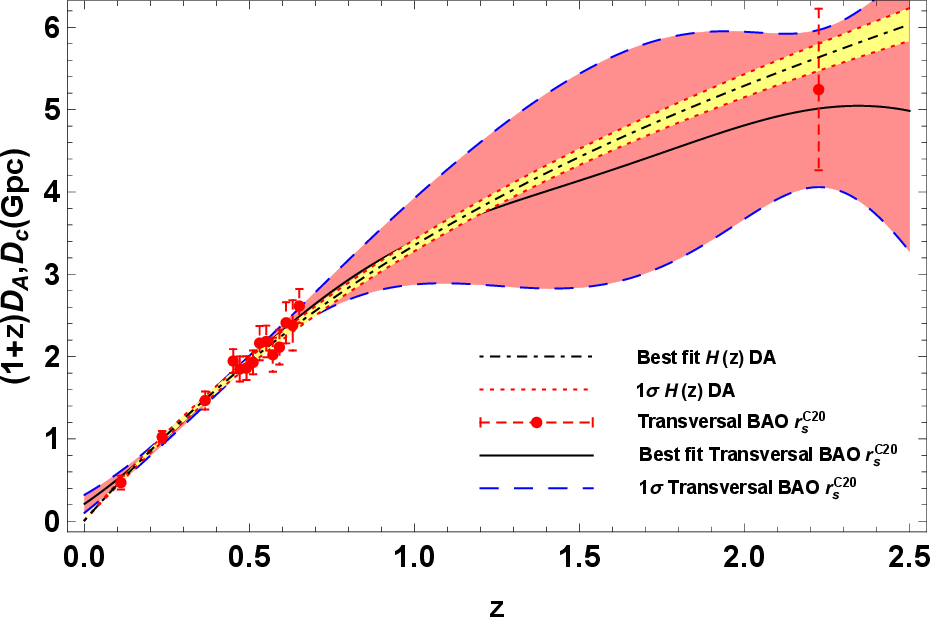}
\includegraphics[width=7cm]{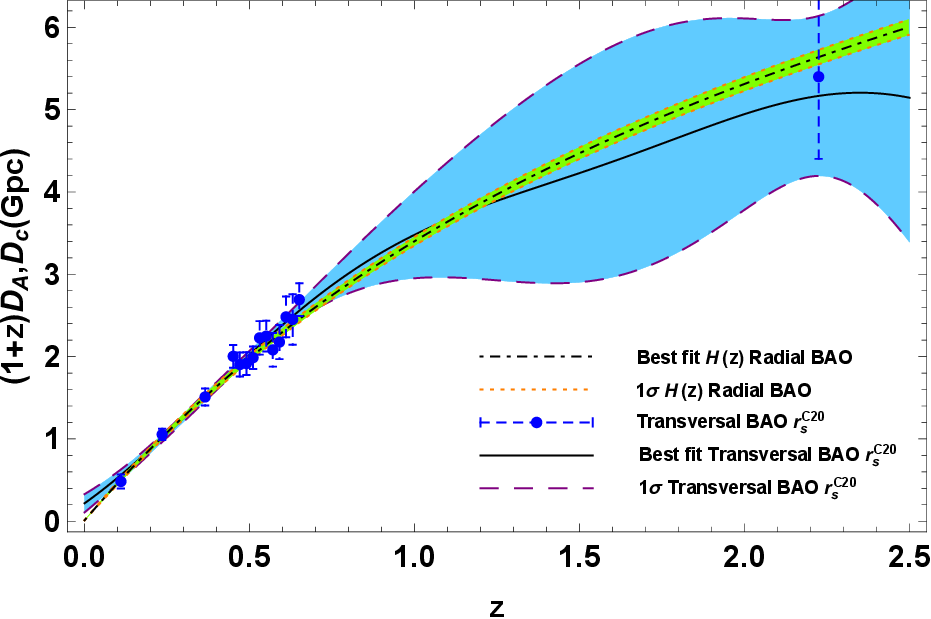}
\includegraphics[width=7cm]{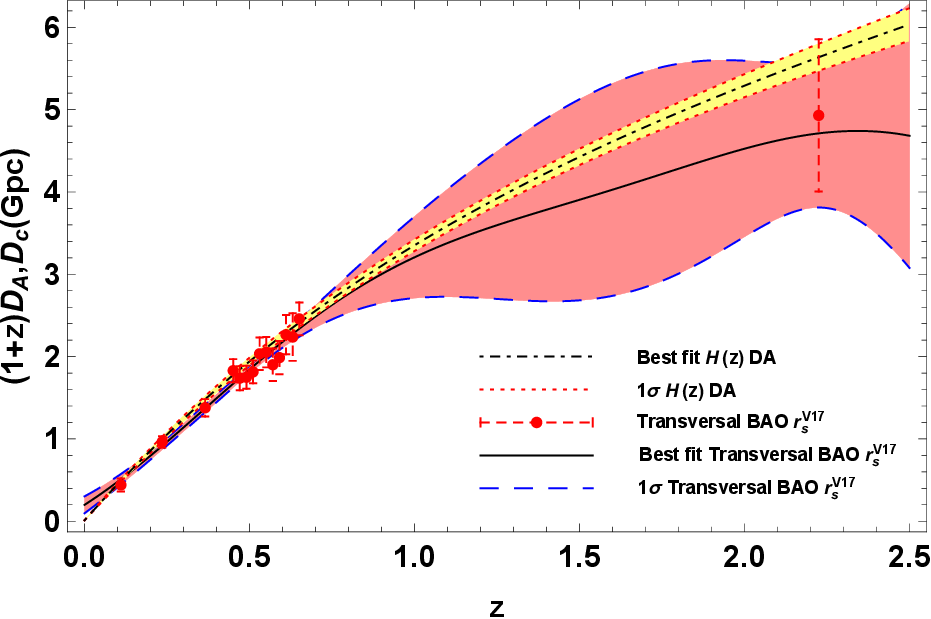}
\includegraphics[width=7cm]{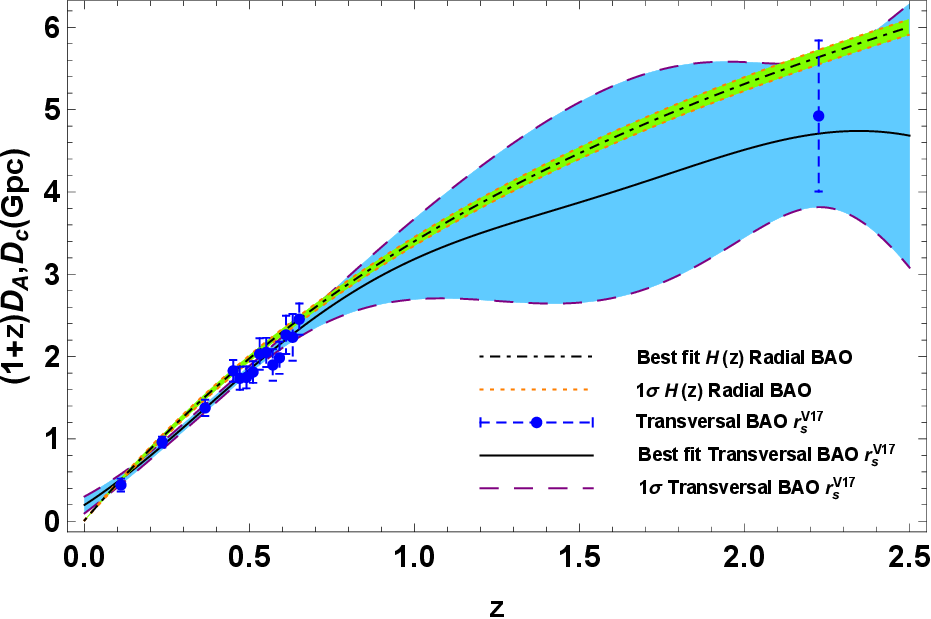}
\caption{\label{fighubble} Reconstructed $H(z)$ (top) and $D_{\rm C}(z)$ (middle and bottom) from Hubble measurements. The sample catalogs of the observed $(1+z)D_{\rm A}(z)$ distributions from the transversal BAO and the corresponding reconstructed curves (middle and bottom)  with the $H_0^{\rm DA}$ (left panel) and $H_0^{\rm RB}$ (right panel) and the priors of $r_{\rm s}^{\rm C20}$ and $r_{\rm s}^{\rm V17}$.}
\end{figure}

Using the Gaussian process, we first obtain the $H(z)$ in 2000 reconstruction bins in the redshift range $0<z<2.5$, and the results are shown in Fig.~\ref{fighubble}. The quantities of the present Hubble parameter $H_0$ from the reconstruction are shown as
\begin{equation}\label{hubb1}
H_0^{\rm DA}={66.3\pm4.22{\rm km \, s^{-1}Mpc^{-1}}}\,,
\end{equation}
\begin{equation}\label{hubb2}
H_0^{\rm RB}={64.4\pm3.56{\rm km \, s^{-1}Mpc^{-1}}}\,,
\end{equation}
here, the $H_0^{\rm DA}$ and $H_0^{\rm RB}$ denote  the present Hubble parameters $H_0$ obtained from the method of differential age and radial BAO observations, respectively. The reconstructed results of $H_0$ are compatible with the observed constraints $H_0={67.0\pm0.9{\rm km\, s^{-1}\,Mpc^{-1}}}$ obtained from the Hubble and SNIa data~\cite{Magana2018}, and are a little less than the results  $H_0={67.2\pm^{1.2}_{1.0}{\rm km\,s^{-1}\,Mpc^{-1}}}$  obtained from  Dark Energy Survey Year 1 clustering combined with BAO and Big Bang Nucleosynthesis~\cite{Abbott2018}.

We used the trapezoidal rule to numerically integrate the reconstructed $H(z)$, and obtain the comoving distance~\cite{holanda2013,Bengaly2022}
\begin{equation}\label{Hzint}
  D_{{\rm C},\,i}=c\int_{0}^{z}{dz'\over{H(z')}}{\approx{c\over{2}}\sum_{i=1}^N{(z_{i+1}-z_i)}{\bigg[{1\over{H_{(z_{i+1})}}}+{1\over{H_{(z_{ i})}}}\bigg]}}\,.
\end{equation}
The uncertainty of the ${D_{{\rm C},\,i}}$ obtained from  reconstruction $H(z_i)$ is given by
\begin{equation}\label{sigDc}
  \sigma_{D_{{\rm C}, \,i}}={c\over{2}}{(z_{i+1}-z_i)}{\bigg[{\sigma^{2}_{H_{i+1}}\over{H^{4}_{i+1}}}+{\sigma^{2}_{H_i}\over{H^{4}_i}}\bigg]}^{1/2}
\end{equation}
The integration in Equ.~{\ref{Hzint}} is performed along  the evenly spaced-out 2000 bins over the redshift range $0<z<2.5$, rather than over the inhomogeneous Hubble $H(z)$ measurements. The reconstructed $D_{\rm C}(z)$ from the observed Hubble $H(z)$ measurements are shown in Fig.~\ref{fighubble}.

To obtain the ADD $D_{\rm A}$ from the transversal BAO data, a sound horizon at radiation drag $r_{\rm s}$  should be determined from the astronomic observations.  The quantity of the sound horizon scale  $r_{\rm s}$  can be calibrated at
$z>1000$ from CMB observations and theoretical assumptions, i.e.,  $r_{\rm s}=147.33\pm 0.49$Mpc and $r_{\rm s}= 152.30\pm 1.3$Mpc from the most recent Planck~\cite{Ade2016} and WMAP9~\cite{Bennett2013} measurements, respectively.
At the relative low redshift, Carvalho, {\it et al}., made constraint on $r_{\rm s}$  by using the SDSS data release 11 galaxies and a prior on the matter density parameter given by the SNIa data (hereafter referred to as $r_{\rm s}^{\rm C20}$)~\cite{Carvalho2020},
\begin{equation}
r_{\rm s}^{\rm C20}={107.4\pm1.7{\,h^{-1}}{\rm Mpc}}\,.
\end{equation}
It should be noted that the measurement of $r_{\rm s}^{\rm C20}$ is model-dependent, since it was estimated by fitting the flat $\Lambda$CDM model to the transverse BAO
data. Here, $h$ can be obtained from the following expression
\begin{equation}
H_0={100 { h}\,{\rm km \,s^{-1} Mpc^{-1}}}\,.
\end{equation}
Verde {\it et al}. also obtained constraints on the length of the low-redshift standard ruler (referred to as $r_{\rm s}^{\rm V17}$)
\begin{equation}
r_{\rm s}^{\rm V17}={101.0\pm2.3{\, h^{-1}}{\rm Mpc}}\,,
\end{equation}
when using SNIa  and BAO measurements~\cite{Verde2017}.
In this work, to avoid the bias caused by the different values of $H_0$ between the Hubble measurements and the transversal BAO observations, we use the corresponding value of $H_0$ reconstructed from the Hubble data and the sound horizon scales $r_{\rm s}^{\rm C20}$ and  $r_{\rm s}^{\rm V17}$ at relative low redshift to  derive the ADD  $D_{\rm A}$.

Then, using the Gaussian process, we reconstruct the ADD $D_{\rm A}(z)$  as a smooth
function of redshift $z$ from the transversal BAO data with the different priors of $r_{\rm s}$ and $H_0$, and the results are shown in Fig.~\ref{fighubble}. From this figure, it can be seen that, at the same redshift, the value of  $D_{\rm A}$ obtained from the  prior $r_{\rm s}^{\rm C20}$ is larger than that obtained from $r_{\rm s}^{\rm V17}$, since  $r_{\rm s}^{\rm C20}>r_{\rm s}^{\rm V17}$ for the same value of  $H_0$.

Following the approach proposed by Refs.~\cite{Maartens2011,Arjona2021}, we can search for the possible deviations from the FLRW universe by comparing the $D_{\rm C}$ obtained from the $H(z)$ measurements with  the $D_{\rm A}$ from transversal BAO observational data in the following expression,
\begin{equation}\label{xiz}
\zeta(z)=1-{\theta_{\rm BAO}\over \theta_{H(z)} }= 1-{D_{\rm C}(z)\over (1+z)D_{\rm A}(z)}\,.
\end{equation}
Here, the observational ADD $D_{\rm A}(z)$ and  comoving distances $D_{\rm C}(z)$ can be obtained from the transversal BAO measurements and the reconstructed $H(z)$ measurements. Any violation from $\zeta(z)=0$ in Eq.~(\ref{xiz}) implies
a non-FLRW universe.
The observed $\zeta_{\rm obs}(z)$ can be obtained from Eq.~(\ref{xiz}), and the corresponding error of $\zeta_{\rm obs}(z)$ is
\begin{equation}
\label{SGL}
\sigma^2_{\zeta_{\rm obs}}={{D_{\rm C}^2\over{(1+z)^2{D_{\rm A}^2}}}}\left[\left({\sigma_{D_{\rm A}(z)}\over{D_{\rm A}(z)}}\right)^2+\left(\sigma_{D_{\rm C}(z)}
\over{D_{\rm C}(z)}\right)^2\right]\,.
\end{equation}
Here, $\sigma_{D_{\rm C}}$ can be obtained from the Hubble parameter with Equ.~\ref{sigDc}. $\sigma_{D_{\rm A}}$ is obtained with
\begin{equation}\label{SigDa}
\sigma^2_{D_{\rm A}}=D_{\rm A}^2\left[\left({\sigma_{r_{\rm s}}\over{r_{\rm s}}}\right)^2+\left(\sigma_{\theta_{\rm BAO}}
\over{\theta_{\rm BAO}}\right)^2\right]\,,
\end{equation}
and variable $\theta$ needs to be converted from degrees to radians.

\begin{figure}[htbp]
\includegraphics[width=7cm]{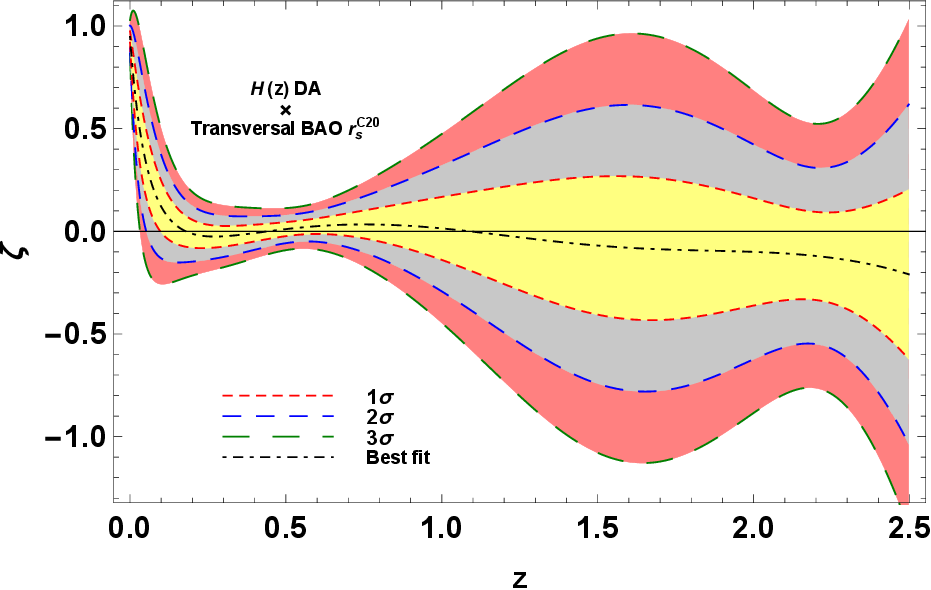}
\includegraphics[width=7cm]{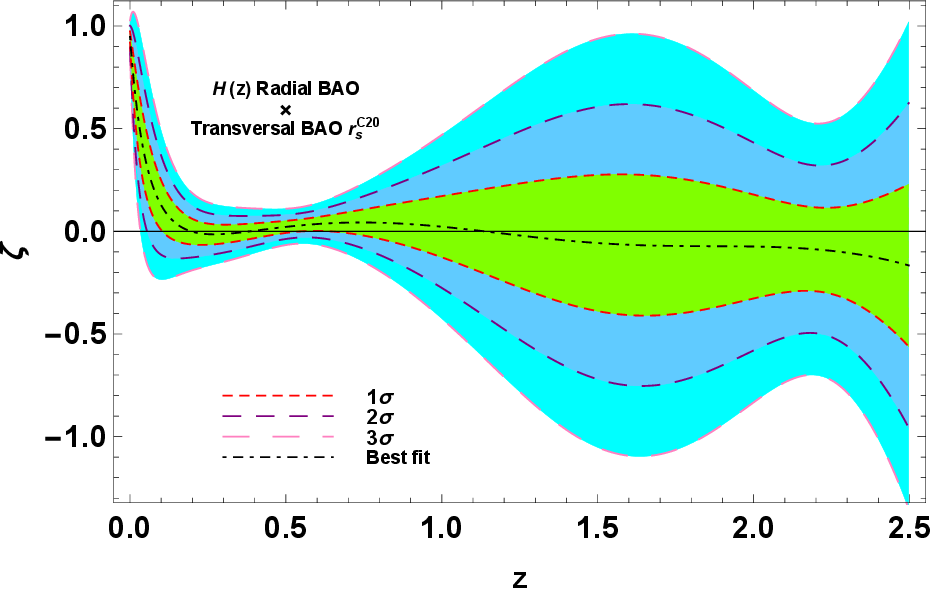}
\includegraphics[width=7cm]{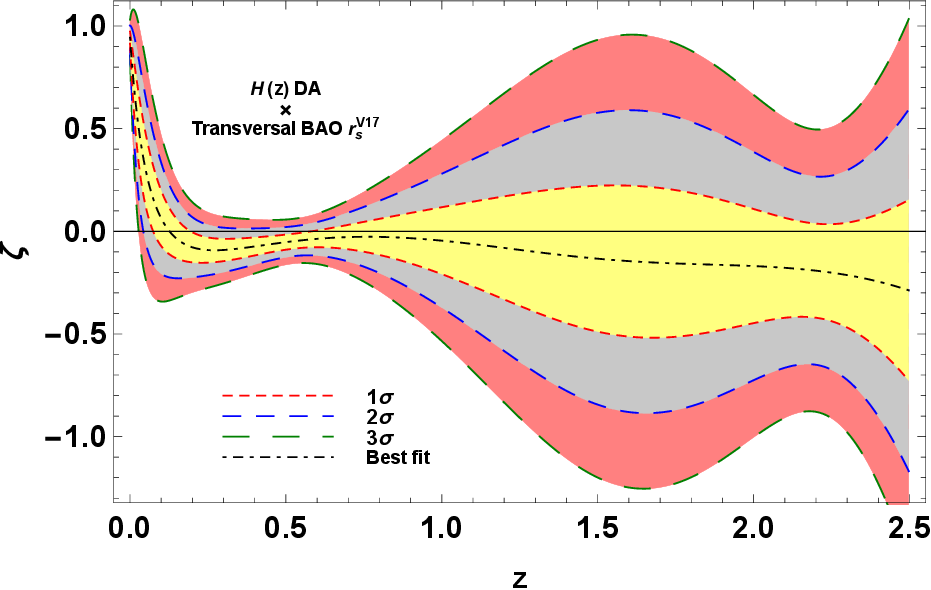}
\includegraphics[width=7cm]{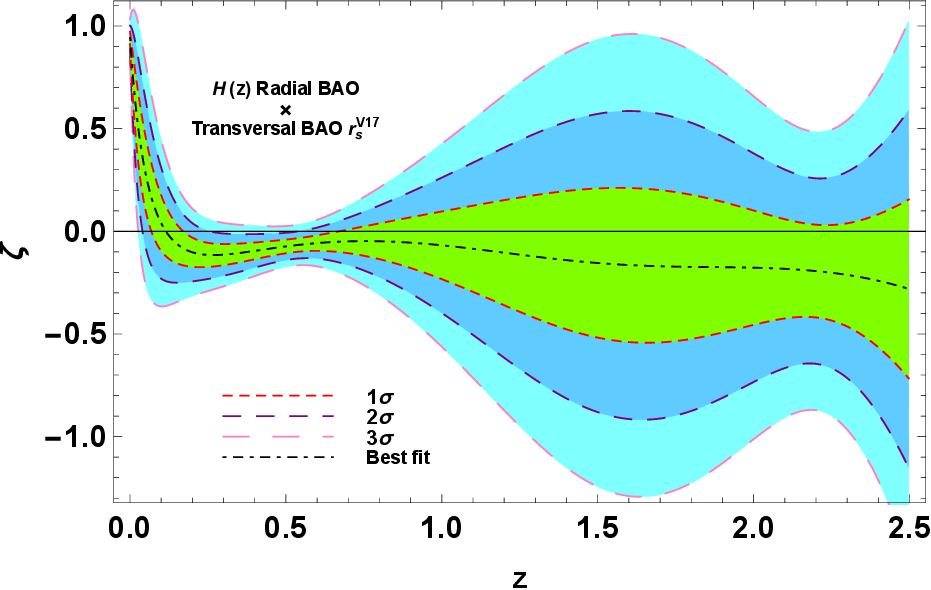}
\caption{\label{kse} { The distribution of deviations from FLRW assumption $\zeta(z)$ obtained from the Hubble measurements of DA method (left panel) and radial BAO measurements (right panel) with the priors $r_{\rm s}^{\rm C20}$ (top) and $r_{\rm s}^{\rm V17}$ (bottom)
obtained in a non-parametric way.} }
\end{figure}

Then we can obtain the function $\zeta(z)$ with a non-parametric method at any redshifts by comparing the reconstructed $D_{\rm A}$ with the reconstructed $D_{\rm C}$ at the same redshift, and show the results  in Fig.~\ref{kse}. 

\begin{figure}[htbp]
\includegraphics[width=8cm]{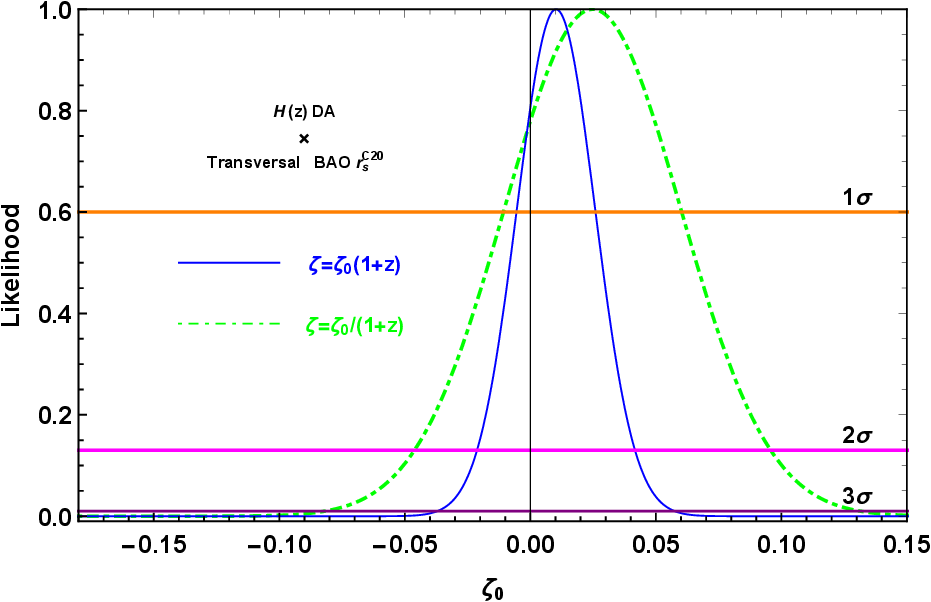}
\includegraphics[width=8cm]{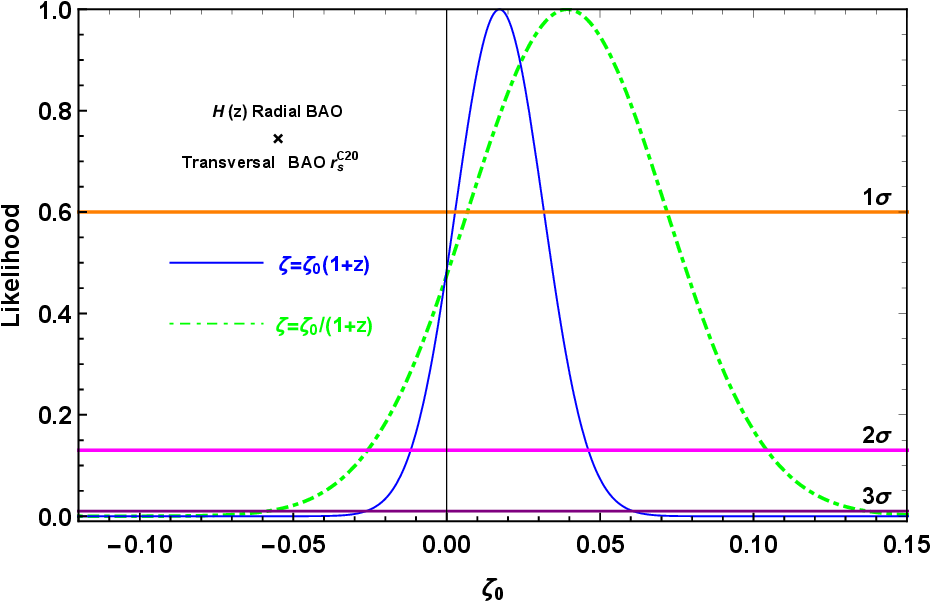}
\includegraphics[width=8cm]{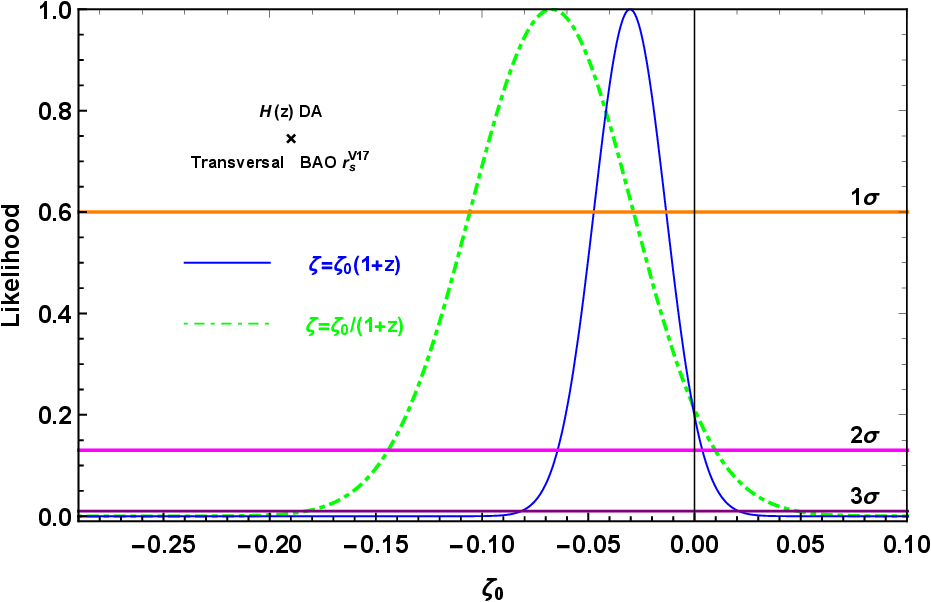}
\includegraphics[width=8cm]{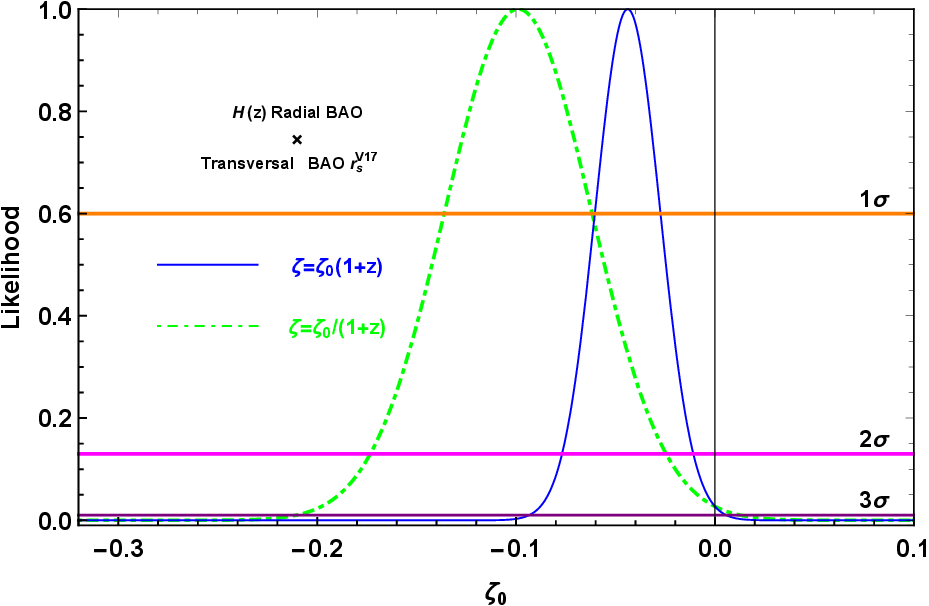}
\caption{\label{likelihood} Likelihood distribution functions obtained from Hubble $H(z)$ measurements of DA method and the transversal BAO observations (left panel) with the priors of $r_{\rm s}^{\rm C20}$ (top) and $r_{\rm s}^{\rm V17}$ (bottom). The same cases are from the  transversal BAO and Hubble $H(z)$ measurements of radial BAO (right panel).}
\end{figure}

As seen from Fig.~\ref{kse}, the best value of a function $\zeta(z)$ is redshift dependent. Therefore, the function $\zeta(z)$  can be parameterized in different forms. It is well known that parameterized method plays an important role in testing the cosmic distance duality relation~\cite{Xubing2022,Fu2019,Li2011,Meng2012,Yang2019} and cosmic opacity~\cite{Liao2016,fuxiangyun20192,fuxiangyun20193}. In our analysis, for  a manageable one-dimensional phase
space and good sensitivity to observational data, we consider the following two parametrization forms. One is inversely proportional to the cosmic scale factor $a=1/(1+z)$,
\begin{equation}\label{xi1}
{\rm P1}:\, \zeta(z)=\zeta_0 (1+z)\,,
\end{equation}
and the other is proportional to the cosmic scale factor $a$,
\begin{equation}\label{xi2}
{\rm P2}:\, \zeta(z)=\zeta_0/(1+z)\,,
\end{equation}
where $\zeta_0$  is a constant parameter and represents possible deviation from the FLRW metric. Then, we can constrain the parameter $\zeta_0$ by comparing the reconstructed comoving distances $D_{\rm C}$ from Hubble $H(z)$ measurements with the angular distances $D_{\rm A}$ from the transversal BAO data at the same redshift. To match the observed $D_{\rm A}$ with the $D_{\rm C}$ at the same redshift, the values of $D_{\rm C}$ are obtained from the reconstructed comoving distances  at the redshift of the transversal BAO measurements. Thus, all 15 available observed data of transversal BAO can be employed to test the FLRW metric.
Therefore, $\chi^2$ is given by
\begin{equation}
\label{chi}
\chi^{2}(\zeta_0)=\sum_i^{N}\frac{{\left[\zeta(z)-
\zeta_{{\rm obs},\,i}(z) \right] }^{2}}{\sigma^2_{\zeta_{{\rm obs},\,i}}}\,.
\end{equation}
Here, $N$ represents the number of the transversal BAO data points, and $N=15$. The free parameter is $\zeta_0$, and the number of the degree of freedom used to perform the fitting procedure is $1$. The reduced $\chi^{2}$ can obtained with $\chi_{\rm red}^{2}=\chi_{\rm min}^{2}/(N-1)$. The constraints on parameter $\zeta_0$ are shown in Fig.~\ref{likelihood} and Table~\ref{likelihood1}.

\begin{table*}[htb]
\label{tab:results}
\begin{center}
\begin{tabular}{ccccc} \hline \hline
& P1: $\zeta_0 {(1+z)}$ & $\chi^2_{\rm red}$ & P2: $\zeta_0 {\frac{1}{1+z}}$ &  $\chi^2_{\rm red}$   \\  [1ex] \hline
$\zeta_0^{\rm {A}\,\diamond}$ & ${0.010{\pm{0.016}}{\pm{0.031}}{\pm{0.049}}}$ & $0.292$ & ${0.025{\pm{0.036}}{\pm{0.071}}{\pm{0.107}}}$ & $0.287$   \\ [1ex]  \hline
$\zeta_0^{\rm {B}\,\diamond}$ & ${0.017{\pm{0.015}}{\pm{0.029}}{\pm{0.044}}}$  & $0.334$ & ${0.039{\pm{0.033}}{\pm{0.066}}{\pm{0.099}}}$ & $0.331$     \\ [1ex]  \hline
$\zeta_0^{\rm {A}\,\star}$ & ${-0.030{\pm{0.017}}{\pm{0.035}}{\pm{0.053}}}$ & $0.272$ & $ {-0.067{\pm{0.039}}{\pm{0.077}}{\pm{0.117}}}$ & $0.281$     \\ [1ex]  \hline
$\zeta_0^{\rm {B}\,\star}$ & ${-0.044{\pm{0.017}}{\pm{0.033}}{\pm{0.051}}}$ & $0.308$ &  $ {-0.099{\pm{0.038}}{\pm{0.075}}{\pm{0.101}}}$ & $0.308$    \\ [1ex]  \hline
$\zeta_0^{\rm {A}\,\dagger}$ & ${-0.001{\pm^{0.075}_{0.091}}{\pm^{0.131}_{0.212}}{\pm^{0.180}_{0.391}}}$ & $0.439$ & ${0.019{\pm^{0.202}_{0.228}}{\pm^{0.346}_{0.569}}{\pm^{0.463}_{1.100}}}$ & $0.439$  \\ [1ex]  \hline
$\zeta_0^{\rm {B}\,\dagger}$ & ${0.011{\pm^{0.066}_{0.085}}{\pm^{0.123}_{0.194}}{\pm^{0.168}_{0.340}}}$ & $0.503$ & ${-0.015{\pm^{0.170}_{0.259}}{\pm^{0.313}_{0.600}}{\pm^{0.432}_{1.110}}}$ & $0.505$   \\
\hline \hline
\end{tabular}
\caption[]{The summary of maximum likelihood estimation results of $\zeta_0$ for the two parametrizations.  The $\zeta_0$ is represented by the best fit value $\zeta_{0,\rm{best}}\pm 1\sigma\pm 2\sigma\pm 3\sigma$ for each dataset. $1\sigma$, $2\sigma$, and $3\sigma$ denote the $68.3\%$, $95.4\%$, and $99.7\%$ CL, respectively. The superscripts A and B represent the cases obtained from Hubble $H(z)$ measurements of the DA method  and the radial BAO observation, respectively. The superscripts $\diamond$, $\star$, and $\dagger$ represent the results obtained with the priors of $r_{\rm s}^{\rm C20}$ (two top lines), the priors of $r_{\rm s}^{\rm V17}$ (two middle lines), and flat priors of $r_{\rm s}$ (two bottom lines), respectively.}
\label{likelihood1}
\end{center}
\end{table*}

\begin{figure}[htbp]
\includegraphics[width=8cm]{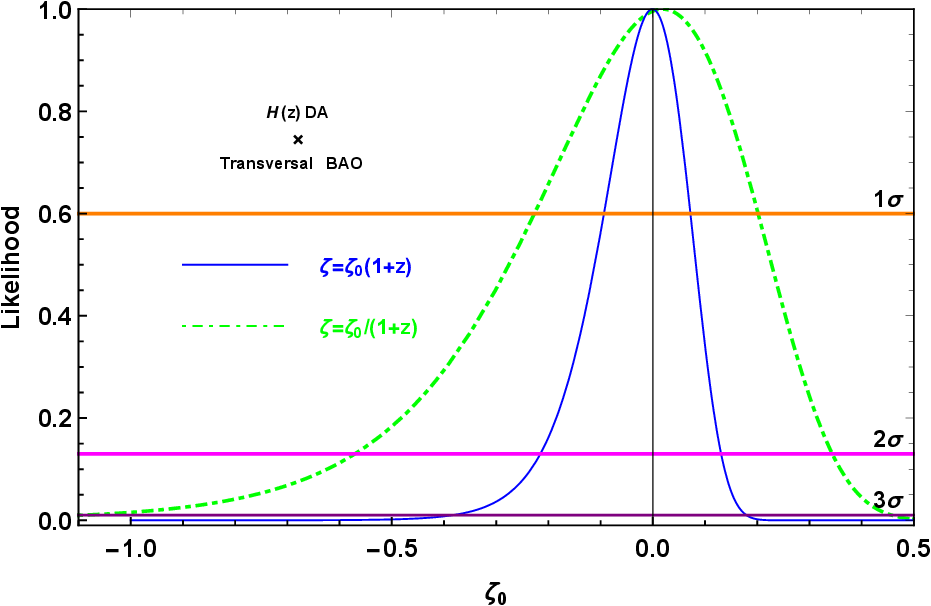}
\includegraphics[width=8cm]{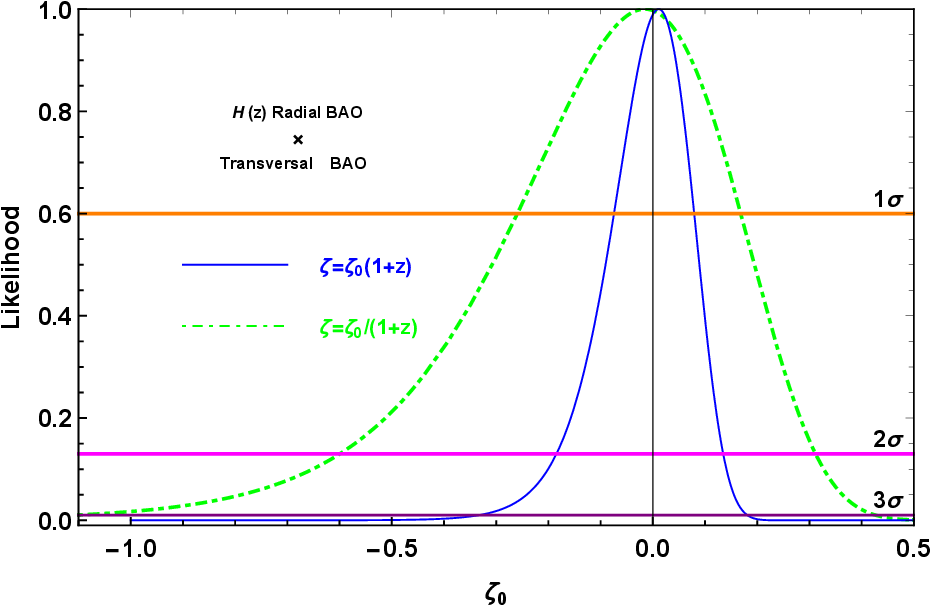}
\caption{\label{LHind} Likelihood distribution functions obtained from Hubble $H(z)$ measurements of DA method and the transversal BAO observations (left panel) with a flat prior of $r_{\rm s}$. The same cases are from the  transversal BAO and Hubble $H(z)$ measurements of radial BAO (right panel). }
\end{figure}

It is obvious to see that the results obtained from the non-parametric and parametric methods are dependent on the priors on the sound horizon scale. To test the FLRW metric independent of the sound horizon scale $r_{\rm s}$, we consider
a fiducial value of $r_{\rm s}$ as a nuisance parameter to determine the ADD $D_{\rm A}$ from transversal BAO measurements, and then marginalize its influence with a flat prior in the analysis. The likelihood distribution $\chi^{\prime\,2}$ can be rewritten as
\begin{equation}
\label{chi2}
\chi^{\prime\,2}(\zeta_0, r_{\rm s})= \sum_i^{N}\frac{{{n_i^2 \over m_i^2}r_{\rm s}^2- 2 {n_i \over m_i}r_{\rm s}+1  }}{\sigma^{\prime\,2}_{{\zeta_{{\rm obs},\,i}}}}\,,
\end{equation}
here, $n_i=1-\zeta(z_i)$, $m_i=\theta_{{\rm BAO},\,i}D_{{\rm C},\,i}$, and
\begin{equation}
\label{sigma0}
\sigma_{\zeta_{{\rm obs},i}}^{\prime\,2}=\left({\sigma_{\theta_{{\rm BAO},i}}\over{\theta_{{\rm BAO},\,i}}}\right)^2+\left(\sigma_{D_{{\rm C},i}(z)}
\over{D_{{\rm C},\,i}(z)}\right)^2\,.
\end{equation}
Then, following the method in Ref.~\cite{Xubing2022,Conley2011}, we marginalize analytically the likelihood
function over $r_{\rm s}$ with the assumption of a flat prior on $r_{\rm s}$, and rewrite the marginalized $\chi_{\rm M}^{\,\prime\,2}$ as
\begin{equation}
\label{chi3}
\chi_{\rm M}^{\prime\,2}(\zeta_0)= C-{B^2\over {A}}+\ln{A\over 2\pi}\,,
\end{equation}
where $A=\sum n_i^2/(m_i^2{\sigma^{\prime\,2}_{{\zeta_{{\rm obs},\,i}}}})$, $B=\sum n_i/(m_i{\sigma^{\prime\,2}_{{\zeta_{{\rm obs},\,i}}}})$, and $C=\sum 1/{\sigma^{\prime\,2}_{{\zeta_{{\rm obs},\,i}}}}$. It should be noted that we adopt the current form of $\zeta(z)$ in Equ.\ref{xiz} (rather than the form in Ref.~\cite{Arjona2021}) to obtain an analytical expression in Equ.\ref{chi3} while marginalizing over $r_{\rm s}$.  The free parameter in this equation is $\zeta_0$, and  the number of degree of freedom is $1$, since the parameter $r_{\rm s}$ has been marginalized. The results are shown in Fig.\ref{LHind} and Table~\ref{likelihood1}.

\section{ Results and analysis}
For the reconstructed results obtained from the deviations of the FLRW metric with the non-parametrization method, seen from Fig.~\ref{kse},  the divergence of $\zeta(z)$ in the redshift range $z<0.1$ may result from the absence of the observed transversal BAO data points. Using the priors of $r_{\rm s}^{\rm C20}$ and $r_{\rm s}^{\rm V17}$, the FLRW metric is consistent with the Hubble $H(z)$ measurements from DA method and the transversal BAO observation at $1\sigma$ and $2\sigma$ confidence level (CL), respectively. And no deviation from the FLRW metric can be found for the Hubble $H(z)$ measurements of radial BAO and the transversal BAO observations at $2\sigma$ and $3\sigma$ CL, respectively.  Our result is
compatible with that from Ref.~\cite{Arjona2021}, in which an isotropic universe was obtained  at $1\sigma$ CL and $2\sigma$ CL from different reconstruction methods and the priors on
the $r_{\rm s}$ and $H_0$. It also shows that the deviation from the FLRW metric is less than that obtained from Ref.~\cite{Bengaly2022}, in which a mild deviation of the FLRW metric was found at $3.5\sigma$ CL  under the assumptions of  $r_{\rm s}^{\rm C20}$ and $H_0=73.4\pm1.4{\rm km\,s^{-1}Mpc^{-1}}$ priors. The greater deviation found in Ref.~\cite{Bengaly2022} is due to the value of $H_0$ being larger than that in our analysis. In Fig.~\ref{kse}, it can  also be seen that the deviation from the FLRW metric obtained from the prior of the sound horizon scale  $r_{\rm s}^{\rm V17}$  is greater than that obtained from the prior $r_{\rm s}^{\rm C20}$, because a larger value of  $r_{\rm s}$ may result in a larger angular distance $D_{\rm A}$, as shown in Sec.~\ref{subsec}. Thus, bias may be caused by the priors of the sound horizon scales in the test of the FLRW metric.
It can be also found that the best-fit value of function $\zeta(z)$ varies with the redshift $z$ regardless of the observational data and the priors on $r_{\rm s}$.

For the constraints on the parameter $\zeta_0$ by comparing the Hubble $H(z)$ measurements from the DA method with the transversal BAO observation, seen from Fig.~\ref{likelihood} and Table~\ref{likelihood1},  we find that the FLRW metric is consistent with the observational data with the priors of $r_{\rm s}^{\rm C20}$ and $r_{\rm s}^{\rm V17}$ at $1\sigma$ and $2\sigma$ CL, respectively. And the FLRW metric is also consistent with the Hubble $H(z)$ measurements from the radial BAO and  transversal BAO observations with the priors of $r_{\rm s}^{\rm C20}$ and $r_{\rm s}^{\rm V17}$ at $2\sigma$ and $3\sigma$ CL, respectively.
For the two observational datasets and  priors of $r_{\rm s}$,
the error bars obtained from parametrization $\rm P1$ are nearly 50\% smaller than those from parametrization $\rm P2$,
although the results are almost independent on the
parametrizations of $\zeta(z)$. It should be noted that the results from the parametrizations are consistent with the results from the non-parametrization method. Thus, the parametric method provides an effective way to check the consistency relation through constraining the parameter $\zeta_0$. In addition, the constraints obtained from the priors of $r_{\rm s}^{\rm V17}$ imply the greater violation from the FLRW metric than that from priors of $r_{\rm s}^{\rm C20}$, which is similar to the results obtained from the non-parametric method. Therefore, for non-parametric and parametric methods, priors of the sound horizon scale will lead to bias in the tests of the FLRW metric, if the exact value of $r_{\rm s}$ cannot be given from astronomical observations.

As for the case of constraints on the parameter $\zeta_0$ with a flat prior of the sound horizon scale $r_{\rm s}$, the FLRW metric is compatible with  the observational data at $1\sigma$ CL, although the error bars are much larger (nearly 5 times)  than those obtained with the  priors of  the specific $r_{\rm s}$. While comparing the two compilations of Hubble measurements with the transversal BAO data, the error bars obtained from parametrization $\rm P1$ are nearly 60\% smaller than those from parametrization $\rm P2$. So, parametrization $\rm P1$ offers a better fit on the observational data. The value of the reduced $\chi^2$ ($\chi^2_{\rm red}$) obtained from the flat prior of $r_{\rm s}$ is closer to $1$ than that from the prior of $r_{\rm s}$.  Thus, the constraint obtained from the flat prior of $r_{\rm s}$ provides the better fit for the observational measurements than that from the priors of $r_{\rm s}$. It should be noted that the value of $\chi^2_{\rm red}$ is below $1$. Therefore, this indicates some overfitting, which may reduce the statistical significance of these results. In addition, the method in our analysis is independent of any cosmological model, except that the radial BAO measurements are obtained under the assumption of $\Lambda$CDM. Therefore, using the Hubble measurements from the differential age method and the transversal BAO measurements, the method for testing the FLRW metric is not only independent of the cosmological model, but also independent of the priors of the values of the sound horizon scale $r_{\rm s}$.

\section{conclusion and discussion}
The CP is one of the two fundamental assumptions of the SCM, and it  is the cornerstone of measuring cosmic distances and clocks through the FLRW metric. It is significant to verify the FLRW metric with new methods and astronomical observations, because any deviations from the FLRW metric may imply that the tension in the SCM might be caused by  an over-simplifying formulation of its fundamental assumptions.

The comoving observed angular modes from the Hubble $H(z)$ measurements and transversal BAO data should be consistent with each other across the expansion history of the Universe, if the space time of our Universe is described by the FLRW metric. Thus, the direct detections of Hubble $H(z)$ measurements and transversal BAO data provide us with the opportunity to test the FLRW metric. In this work, following the non-parametric method in Refs.~\cite{Arjona2021,Bengaly2022}, we first test the FLRW metric by comparing the Hubble $H(z)$ measurements obtained from the differential age method (or determination of the
BAO peak in the radial direction)  with transversal BAO measurements. The function $\zeta(z)=1-\theta_{\rm BAO}/{\theta_{H(z)}}$ is adopted  to probe the possible deviations from the FLRW metric. We use the best-fitted sound horizon scale $r_{\rm s}$ from the transversal BAO mode at the low redshift and the value of $H_0$  reconstructed  from the Hubble measurements  to derive the ADDs. The function $\zeta(z)$ at any redshift are obtained  with the Gaussian process. The results show that FLRW metric is consistent with the observations regardless of the priors of the sound horizon scale, and the observational data favors a function $\zeta(z)$ varying with redshift. It can be also concluded that the deviation from the FLRW metric is dependent on the priors on the sound horizon scale $r_{\rm s}$, since a smaller value of  $r_{\rm s}$ may result in a smaller ADD.

Then, we employ two parametrizations to describe the function $\zeta(z)$ that evolves with the redshift $z$,  namely, $\zeta(z)=\zeta_0 (1+z)$ ($\rm P1$) and $\zeta(z)=\zeta_0/(1+z)$ ($\rm P2$), and test the FLRW metric by constraining the parameter $\zeta_0$ under the priors of sound horizon scale. Our results show that $\rm P1$ offers a much more rigorous constraint on the parameter $\zeta_0 $ than $\rm P2$. Compared with the results from the non-parametric method,  the same results are obtained from the parametric method for the observational data and the priors of the sound horizon scale $r_{\rm s}$. Therefore, the parametric method offers an effective way to test the FLRW metric by constraining the parameter $\zeta_0$. The results also imply that in the test of the  FLRW metric, some bias might be caused by the priors of the sound horizon scale $r_{\rm s}$ when the exact value of observation $r_{\rm s}$ is not determined.

To test the FLRW metric independent of the sound horizon scale $r_{\rm s}$, we consider
a fiducial value of $r_{\rm s}$ as a nuisance parameter to determine the ADD $D_{\rm A}$ from transversal BAO data, and then marginalize its influence with a flat prior in the analysis. Results show that the FLRW metric is compatible with  the observational data at $1\sigma$ CL, although the ability to constrain  the parameter $\zeta_0$ is weaker than that obtained from  the priors of the sound horizon scale $r_{\rm s}$. Furthermore,  the method to test the FLRW metric with the Hubble measurements from the differential age method is independent of any cosmological model. Due to the limited  observed data available and its large error at present, the ability to constrain  the parameter $\zeta_0$ is weak in this work. With the developments of powerful optical and
radio telescopes, we can better measure the Hubble parameter using the cosmic chronometer and BAO methods. In addition, the neutral hydrogen intensity mapping technique
can be used to measure the BAO signals more efficiently, and 200 observational data at $0<z<2.5$ will be realized in the coming decades~\cite{Zhang2020}. So, in the following work, simulation data  can be used to detect the ability of future observational data to test the FLRW metric. As the quality and quantity of measurements for future Hubble and BAO measurements increase, the parametric method in our analysis
will be a powerful way to test the consistency relation both independent of the cosmological model and the sound horizon scale.

\begin{acknowledgments}

We very much appreciate helpful comments and suggestions from anonymous referees, and helpful discussion from Hongwei Yu, Puxun Wu, and Zhengxiang Li. This work was supported by the National Natural Science Foundation of China under
Grants No. 12375045, No. 12305056, No. 12105097 and No. 12205093; the Hunan
Provincial Natural Science Foundation of China under Grants No.
12JJA001 and No. 2020JJ4284; the Natural Science Research Project of Education Department of Anhui Province No. 2022AH051634, the Science Research Fund of Hunan Provincial Education Department No. 21A0297.

\end{acknowledgments}

\end{document}